\documentclass[preprint]{jpsj2}

\title{Magnetic Properties of 2-Dimensional Dipolar Squares: Boundary Geometry Dependence}
\usepackage{graphicx, color}
\def\nle{\ \raise.3ex\hbox{$<$}\kern-0.8em\lower.7ex\hbox{$\sim$}\ }
\def\nge{\ \raise.3ex\hbox{$>$}\kern-0.8em\lower.7ex\hbox{$\sim$}\ }

\def\Phip4{\Phi=\pi/4}

\author{Ryoko \textsc{Sugano}$^{1}$\thanks{E-mail address:
sugano@rd.hitachi.co.jp}, 
Katsuyoshi \textsc{Matsushita}$^{2}$\thanks{E-mail address:
kmatsu@issp.u-tokyo.ac.jp}, 
Akiyoshi \textsc{Kuroda}$^{2}$\thanks{E-mail address: kro@issp.u-tokyo.ac.jp},
Yusuke \textsc{Tomita}$^{2}$\thanks{E-mail address:
ytomita@issp.u-tokyo.ac.jp} and 
Hajime \textsc{Takayama}$^{2}$\thanks{E-mail address:
takayama@issp.u-tokyo.ac.jp}} 
\inst{
$^{1}$ Advanced Research Laboratory, Hitachi, Ltd.,
   1-280 Higashi-Koigakubo, Kokubunji-shi, Tokyo 185-8601,
   Japan.\\$^{2}$ Institute for Solid State Physics, University of
   Tokyo, 5-1-5 Kashiwanoha, Kashiwa, Chiba 277-8581, Japan.}
\recdate{\today}
\abst{
By means of the molecular dynamics simulation on gradual cooling processes, we investigate magnetic properties of classical spin systems only with the magnetic dipole-dipole interaction, which we call dipolar systems.
Focusing on their finite-size effect, particularly their boundary geometry dependence, we study two finite dipolar squares cut out from a square lattice with $\Phi=0$ and $\pi/4$, where $\Phi$ is an angle between the direction of the lattice axis and that of the square boundary.
Distinctly different results are obtained in the two dipolar squares. 
In the $\Phi=0$ square, the ``from-edge-to-interior freezing'' of spins is observed. 
Its ground state has a multi-domain structure whose domains consist of the two among infinitely (continuously) degenerated Luttinger-Tisza (LT) ground-state orders on a bulk square lattice, i.e., the two antiferromagnetically aligned ferromagnetic chains (af-FMC) orders directed in parallel to the two lattice axes.
In the $\Phi=\pi/4$ square, on the other hand, the freezing starts from the interior of the square, and its ground state is nearly in a single domain with one of the two af-FMC orders.
These geometry effects are argued to originate from the anisotropic nature of the dipole-dipole interaction which depends on the relative direction of sites in a real space of the interacting spins.
}

\begin{document}
\sloppy
\maketitle

\section{Introduction}

In recent years, systems consisting of arrayed single-domain ferromagnetic nanoparticles have attracted much attention as a possible element with high storage density.~\cite{Martin}
They exhibit rich variety of magnetic phenomena caused by the interplay between the dipole-dipole interactions and the magnetic anisotropic energy of magnetic moments of nanoparticles.~\cite{DeBell00} 
There also appeared the works on finite-size effects on such systems.~\cite{Stamps,Denisov,Kayali,KSKTT,Takagaki}
In order to get a deeper insight into these rather new and complicated phenomena, we have been studying magnetic properties of the systems of magnetic moments between which only the dipole-dipole interaction is present, which we call dipolar systems. 
Particularly, we have focused on dipolar systems of a finite size, and have found a peculiar ``from-edge-to-interior freezing'' in a dipolar cube by a finite-temperature molecular dynamics (MD) simulation.~\cite{KSKTT}
As the temperature is decreased, magnetic moments on each edge of the cube predominantly start to freeze and then domains with short-raged orders corresponding to the bulk Luttinger-Tisza (LT) orders~\cite{LT} grow from edges to the interior.
This phenomenon is a very novel one, which, to our knowledges, has not been observed in uniform magnetic systems with ordinary short-ranged exchange interactions.

The magnetic dipole-dipole interaction is quite ubiquitous and has been well known for long.
It is long-ranged, i.e., the strength of the interaction between magnetic moments (which we call simply as spins hereafter) with relative distance $r$ is proportional to $r^{-3}$. 
It has more peculiar nature though not so emphasized in general. 
Namely, not only the strength but also the sign of the interaction depends on the direction of relative positions in a real space of the interacting spins. 
This anisotropic nature intrinsically gives rise to the frustration effect even if spins lie periodically on a certain lattice, and it yields different ground-state magnetic orders depending on the lattice structure as pointed out by LT more than a half century ago.~\cite{LT} 
[Recently, the lower energy states than the LT ground state have been found for the BCC dipolar lattice.~\cite{Tomita-etal}]
We can then naturally expect that this anisotropic nature, combined with the geometry (boundary) effect, plays a key role on determining the magnetic properties of finite dipolar systems.
Actually, a variety of phenomena in various finite dipolar systems have been found in our preliminary study.~\cite{ours-new}  
There have appeared, on the other hand, many numerical works which point out the important role of the dipole-dipole interaction on the magnetic properties of an individual ferromagnet where the ferromagnetic exchange interaction dominates the dipole-dipole interaction.~\cite{DeBell00,Chui,Sasaki1,MacIsaac,Sasaki2,Iglesias,Endoh,Sasaki3,Cannas}
However, the proper understanding of genuine dipolar systems of a finite size is quite interesting by itself, and is also considered to be one of the fundamental bases for technological development of nanomagnetism.

\begin{figure}[b] 
\begin{center} 
\includegraphics[width=0.8\linewidth]{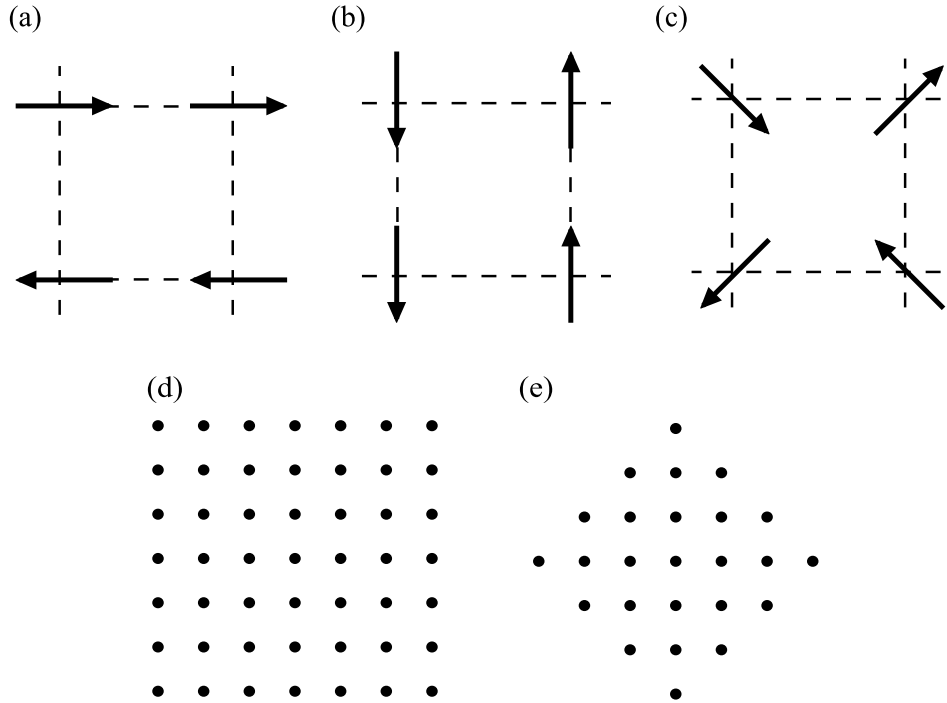} 
\end{center} 
\caption{Spin configurations in the ground state of the bulk dipolar square lattice (a-c). 
They have the 2 $\times$ 2 periodicity, i.e., the LT order which have the continuous $O(2)$ degeneracy: $\Theta = 0$ (a), $\pi/2$ (b), and  $\pi/4$ (c). 
Dipolar squares cut out from the lattice with $\Phi=0$ and $\Phip4$ are shown in (d) and (e), respectively.
} 
\label{GS} 
\end{figure}

In the present paper, we study the ground-state and the freezing characteristics of finite dipolar squares cut out from a square lattice, which are the simplest finite dipolar systems.
The corresponding bulk (infinite) dipolar square lattice has the continuous $O$(2) degeneracy in its ground state~\cite{BGI,DBell}. 
The ground-state order is a 2-dimensional (2D) version of the LT order on a simple cubic lattice.~\cite{LT} 
It is characterized by the order parameter ${\cal M}$ and the state parameter $\Theta$, where ${\cal M}$ is the magnitude of the order parameter vector ${\vec M}$ whose $\alpha$-th component is defined by  
\begin{equation} 
M_{\alpha} = \sum_i (-1)^{\sum_{\beta \not = \alpha} r_i^{\beta}}S^{\alpha}_{i}, 
\label{eq:def-M} 
\end{equation} 
and $\Theta$ is defined by  
\begin{equation}
\Theta=\arctan(\frac{M_y}{M_x}).
\label{eq:def-Theta}
\end{equation} 
In Eq.~(\ref{eq:def-M}), $S_i^{\alpha}$ and $r_i^{\alpha}$ are respectively the $\alpha$-th component and the coordinate in unit of the lattice distance of the $i$-th classical Heisenberg spin.  
The definition of Eq.~(\ref{eq:def-Theta}) is introduced by taking into account the fact that, in the ground state, all the spins are in parallel with the lattice plane which is specified as $x$-$y$ plain.  
Three examples of the LT order are shown in Figs.~\ref{GS}(a-c). 
We call the LT orders with $\Theta=0$ (a) and $\pi/2$ (b) the antiferromagnetically aligned ferromagnetic chains (af-FMC) order, and those with $\Theta=\pi/4$ (c) and $3\pi/4$ as the micro-vortex order.

\begin{figure}[t] 
\begin{center} 
\includegraphics[width=0.8\linewidth]{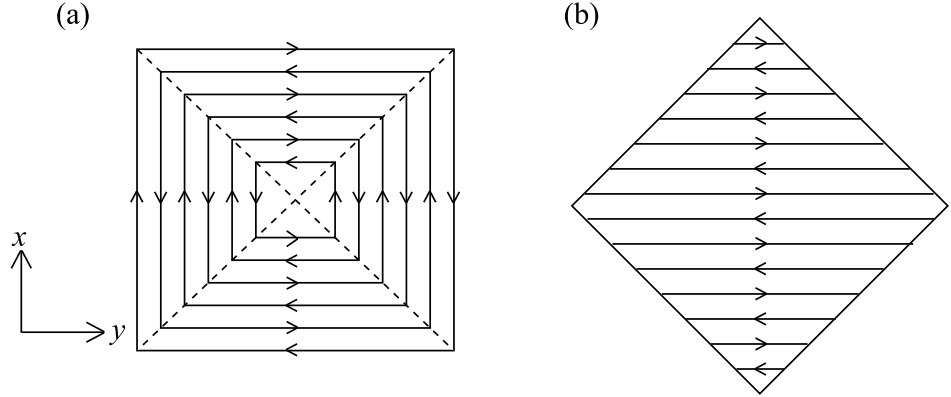} 
\end{center} 
\caption{Schematic pictures of spin configurations in the ground state for $\Phi = 0$ (a), $\pi/4$ (b). The arrows denote the spin direction.} 
\label{gr-st-scheme} 
\end{figure}

A main purpose of the present work is to clarify the role of the unique anisotropic nature of the dipole-dipole interaction mentioned above on determining magnetic properties of square dipolar systems of a finite size.
For this purpose, with focusing on their boundary-geometry dependence, we study two types of square cut out from a square lattice with $\Phi=0$ and $\pi/4$ with $\Phi$ being an angle between an edge of the square and the lattice axis (see Fig.~\ref{GS}(d) and (e)).
By a similar MD simulation to the one we performed previously,\cite{KSKTT} we have found significantly different ground states in the two dipolar squares as schematically shown in Fig.~\ref{gr-st-scheme}.

The ground state of the $\Phi = 0$ dipolar square (Fig.~\ref{gr-st-scheme}(a)) consists of four domains with the two af-FMC orders shown in Figs.~\ref{GS}(a) and (b).
The spins of each af-FMC order align in parallel to the corresponding lattice axis. 
This af-FMC order is attributed to the symmetry reduction of the spin Hamiltonian from global $O(2)$ in the corresponding bulk system to Z$_2$ on four edges of the square.  
The from-edge-to-interior freezing starts nearly from a temperature, denoted as $T^*$, where the LT short-range order becomes comparable with the linear dimension of the system $L$.
These results are similar to, and naturally expected from our previous results on the dipolar cube.~\cite{KSKTT}
The ground state of the $\Phi = \pi/4$ dipolar square (Fig.~\ref{gr-st-scheme}(b)), on the other hand, consists of a single domain with small modulations started from the top and bottom corners (not shown in the schematic figure). 
Each domain has an af-FMC order along one of the diagonals of the square, which is one of the lattice axes. 
This single domain configuration comes out from an almost complete disappearance of the boundary effect from the zig-zag edges. 
The symmetry reduction to Z$_2$ occurs only around each corner and it stabilizes the ground state. 
The difference of the symmetry reduction on edges between the two squares is reflected in their freezing characteristics which differ significantly from each other.

The present paper is organized as follows. 
In \S 2 we explain our model and numerical method. 
After pointing out common thermodynamic behavior of the two dipolar squares 
in \S 3.1, we discuss properties of the ground state and a few low lying states and the freezing characteristics of the $\Phi=0$ and $\Phi=\pi/4$ dipolar squares in \S 3.2 and \S 3.3, respectively. 
Section 4 is devoted to summary and discussion of the present work.

\section{Model and Method}\setcounter{figure}{0}\label{MM}

The model we study consists of classical Heisenberg spins, $\{{\vec S}_i\}$, which are arrayed in a finite square. 
They correspond to magnetic moments of single-domain ferromagnetic particles normalized by magnetization $M_{\rm s}$ of the latter. 
The dipole-dipole interaction energy between spins is written as
\begin{eqnarray}
{\cal H} = \frac{J}{2a^3}\sum_{<i,j>} {\vec S}_i \cdot \frac{1-3{\vec e}_{ij} \otimes {\vec e}_{ij} }{r_{ij}^3} \cdot {\vec S}_j. \label{Ham}
\end{eqnarray}
Here, ${r}_{ij}$ denotes the dimensionless length between sites $i$ and $j$ normalized by the lattice constant, $a$, and ${\vec e}_{ij}$ the unit vector along the direction from site $i$ to site $j$. 
The value of $M_s^2$ is contained in the coupling constant, $J$, and $J/a^3$ is set as the energy (and temperature with $k_{\rm B}=1$) unit of the present model. 
For example, in a Ni nanoparticle array with a 30 nm particle diameter and $a\simeq 100$~nm~\cite{Fuzi04}, $J/a^3$ is estimated to be about 5 K. 
If Co is replaced for Ni, $J/a^3$ is expected to become of the order of several tens of K.

The freezing characteristics of the model are analyzed using the Landau-Lifshitz-Gilbert (LLG) equation~\cite{LLG}, 
\begin{eqnarray}
\frac{d{\vec S_i}}{dt} = \frac{\gamma}{1+\alpha^2} {\vec S_i} \times \left\{ {\vec H_{{\rm eff},i}} + \frac{\alpha}{M_{\rm s}} {\vec S_i} \times {\vec H_{{\rm eff},i}} \right\}, \label{LLG}
\end{eqnarray}
with the effective magnetic field, ${\vec H_{{\rm eff},i}}$, given by
\begin{eqnarray}
{\vec H_{{\rm eff},i}} = - \frac{1}{M_{\rm s}}\frac{\partial{\cal H}}{\partial\vec{S}_i} + {\vec f}_i,
\label{eq:H-eff}
\end{eqnarray} 
where ${\cal H}$ is the dipole-dipole interaction of Eq.~(\ref{Ham}).
To include the heat-bath effect of temperature $T$, we introduce random forces $\{{\vec f}_i\}$ which obey the distribution laws, 
\begin{equation}
\left<{\vec f}_i\right> = {\vec 0}, \ \ \ \left<{\vec f}_i(t) \otimes {\vec f}_j(0)\right> = {1 \over \gamma}T\alpha M_s\delta_{ij}\delta(t){1},
\label{eq:randomF}
\end{equation}
where $\alpha$ and $\gamma$ are the so-called Gilbert damping constant independent of $T$ and the gyromagnetic constant, respectively.

In the present study, we solve the above set of equations by using the Euler scheme.
The typical Larmor period, $a^3M_{\rm s}/\gamma J$, multiplied by $(1+\alpha^2)$, is used as a time unit. 
We further set the damping constant $\alpha$ to 0.18, and the time step of integrating Eq.~(\ref{LLG}), $\Delta t$, to 0.0088. 
With these values of the parameters, the Larmor precession of a spin due to the internal field of averaged magnitude damps in a time comparable to one period of the precession ($\sim 200 \Delta t$) at $T=0$.
We perform typically 8 cooling runs with a fixed cooling rate, in which we use different sets of random numbers generating their initial configurations as well as fluctuation forces. 
For each cooling run, the temperature is initially set to about 1 ($=J/a^3$) and is decreased by a step of $\Delta T = 0.0015$. 
At each temperature, Eq.~(\ref{LLG}) is integrated over a period of $\tau=$ 6 $\times$ 10$^4$ $\Delta t$.
The quantities of interest $Q$ at each temperature are time averaged over the whole period of $\tau$, and are denoted by $\langle Q \rangle$. 
We regard it a quasi-equilibrium average of $Q$ in a finite dipolar system, postponing the reasoning of our chosen value of $\tau$ to \S 4.
The obtained averages are almost independent of the cooling process at high temperatures but not necessarily at low temperatures due to the intrinsic frustration effect of the dipole-dipole interactions.
We call an average of $\langle Q \rangle$ over different cooling runs as the thermal average and denote it as $[\langle Q \rangle]_{\rm r}$.

In order to investigate the boundary effect, we carry out simulations on $\Phi=0$ dipolar squares whose linear size along an edge, $L_e$, are set to 15 and 16, and on $\Phip4$ dipolar squares whose linear size along a diagonal direction, $L_d$, are set to 22 and 23. 
The total number of sites, $N$, of these squares are comparable: $N=225,\ 256$ for $\Phi=0$ respectively with $L_e=15,\ 16$, and $N=264,\ 265$ 
for $\Phip4$ respectively with $L_d=22,\ 23$. 
The $\Phi=0$ square with $L_e=15$ is enough large for the from-edge-to-interior freezing to be observed.

Before going into detailed discussions of the simulated results, we note here that the magnitudes of fluctuations, $\langle (S_i^\alpha - \langle S_i^\alpha \rangle )^2 \rangle$, are rather isotropic, though spins lie almost within a plane of the square on average, i.e., $\langle S_i^z \rangle \simeq 0$ at all temperatures we have examined. 
In the present work we do not explicitly examine roles of fluctuation of $S_i^z$, or in other words, differences between magnetic properties of the Heisenberg and XY dipolar models, leaving them as a future problem.

\section{Results}\setcounter{figure}{0}\label{Re}

\begin{figure}[t]
\begin{center}
\includegraphics[width=0.6\linewidth]{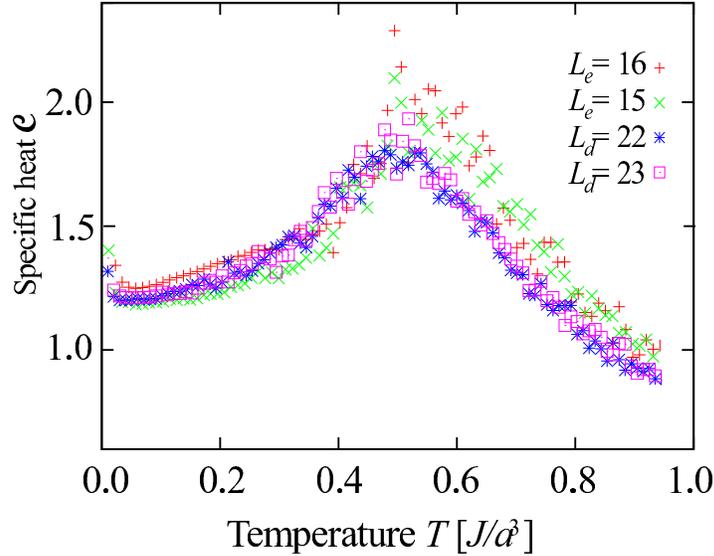}
\end{center}
\caption{Specific heat ${\cal C}$ of the dipolar squares as a function of temperature.}\label{SpeciC}
\end{figure}

\begin{figure}
\begin{center}
\includegraphics[width=0.6\linewidth]{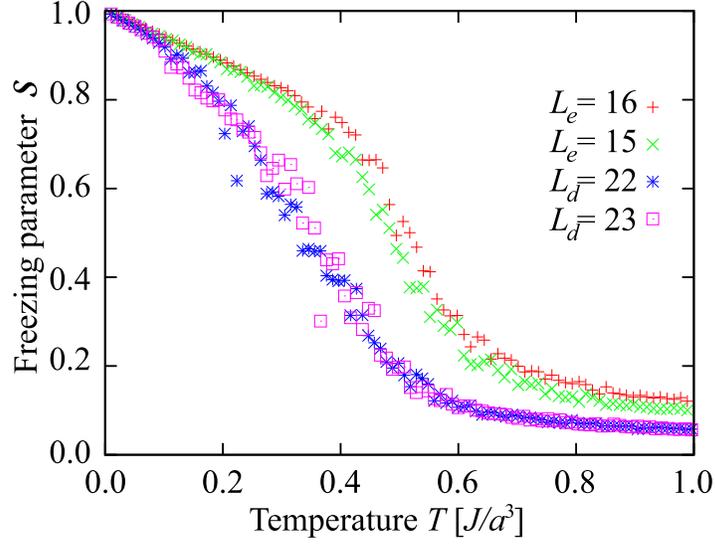}
\end{center}
\caption{The magnitudes of freezing parameter ${\cal S}$ as a function of $T$.}\label{PV}
\end{figure}

\subsection{Thermodynamic quantities}\label{tq}

Let us start our discussion with the thermal average of the specific heat, ${\cal C}$, obtained by the present simulation. 
It is shown in Fig.~\ref{SpeciC}.
Although the data of $\Phi=0$ dipolar squares are rather fluctuating, the four sets of data behave similarly and exhibit a peak at temperature $T \equiv T^* \simeq 0.50 \sim 0.55$. 
The latter is interpreted as a temperature, at which the averaged length scale of the LT short-ranged orders becomes comparable to the system size.
We may say that ${\cal C}$ as well as $T^*$ do not depend sensitively on the boundary geometry. 
The freezing parameter ${\cal S}$, on the other hand, behaves differently in the $\Phi=0$ and $\Phip4$ dipolar squares as seen in Fig.~\ref{PV}.
Here we define ${\cal S}$ as
\begin{eqnarray}
 {\cal S} = \frac{1}{N} \sum_i [ {\cal S}_i ]_{\rm r},\label{freezingS}\\
 {\cal S}_i = |\langle {\vec S_i} \rangle|\ \ {\rm with} \ \ 
\langle {\vec S_i} \rangle = \frac{1}{\tau} \int_\tau \vec{S}_i(t)dt.
\label{Si}
\end{eqnarray}
Concerned with the order parameter ${\cal M}$ similarly defined as 
\begin{equation}
  {\cal M} = [ M_{\rm LT}]_{\rm r}\ \ {\rm with} \ \ M_{\rm LT} =\frac{1}{N} |\langle{\vec M}\rangle|,\label{LT-orderP}
\end{equation}
with ${\vec M}$ being the order parameter vector of Eq.~(\ref{eq:def-M}), $M_{\rm LT}$ turns out to depend not only on the geometry of the squares but also on a cooling run for a fixed geometry (see Fig.~\ref{MLT-0} below).   
These results imply the difference in the freezing characteristics and so the properties of the ground state as well as low-lying excited states of the two dipolar squares whose details we discuss separately below.

\subsection{$\Phi=0$ dipolar squares}\label{3.2}

\begin{figure} [t]
\begin{center}
\includegraphics[width=0.6\linewidth]{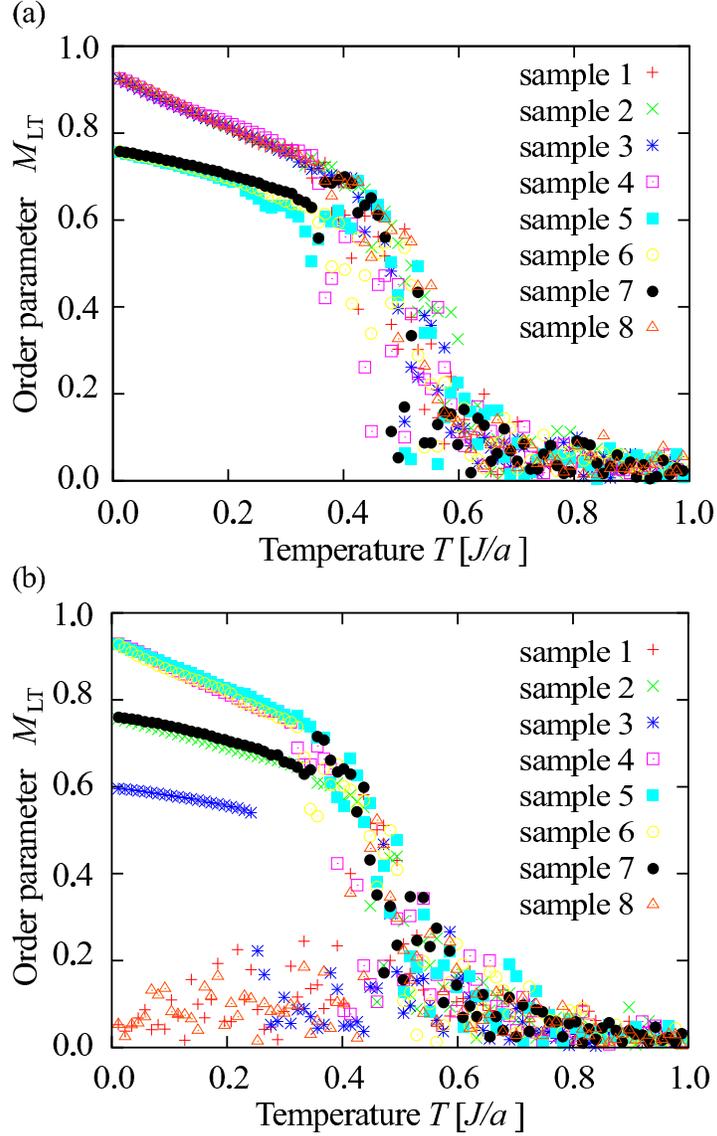}
\end{center}
\caption{The order parameter $M_{\rm LT}$ of $\Phi=0$ squares in each cooling run: (a) $L_e=16$ and (b) $L_e=15$.}
\label{MLT-0}
\end{figure}

\begin{figure}
\begin{center}
\includegraphics[width=0.6\linewidth]{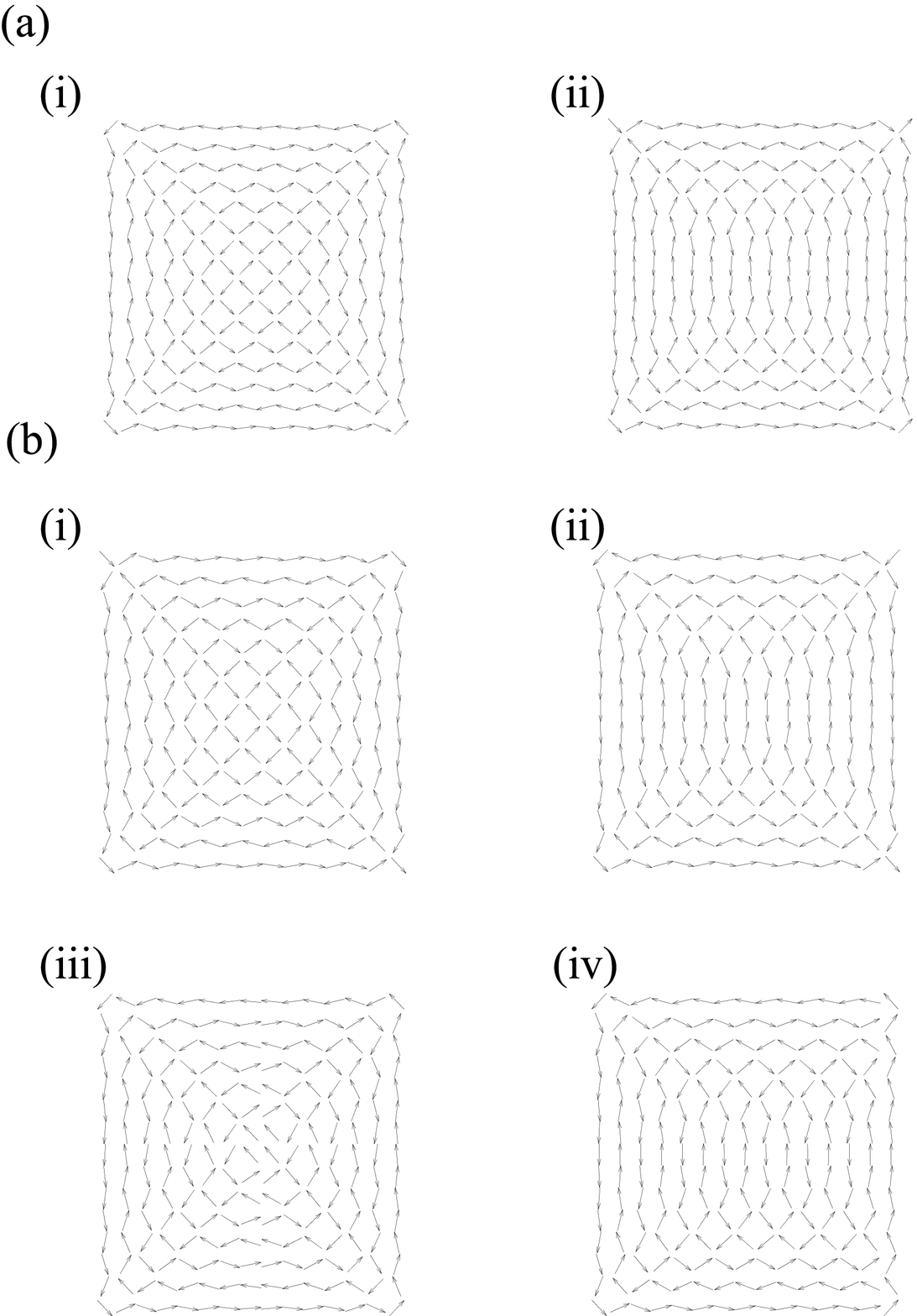}
\end{center}
\caption{Spin configurations at the lowest temperature obtained by the  cooling processes shown in Fig.~\ref{MLT-0}: (a) $L_e=16$ and (b) $L_e=15$}
\label{gr-st-0}
\end{figure}

We show the order parameter $M_{\rm LT}$ of these dipolar squares observed in eight independent cooling runs in Fig.~\ref{MLT-0}.
For the $L_e=15,\ (16)$ square there appear four (two) branches at low temperatures. 
The spin configurations at the lowest temperature for these branches are shown in Fig.~\ref{gr-st-0}. 
The ground-state configuration of the $L_e=16$ (even $L_e$) square shown in  Fig.~\ref{gr-st-0}(a-i) corresponds to the schematic one shown in Fig.~\ref{gr-st-scheme}(a): spins on four edges and on four corners draw a closed loop. 
An almost identical pattern appears on a surface of the dipolar cube examined before~\cite{KSKTT}. 
In the excited state shown in Fig.~\ref{gr-st-0}(a-ii), the direction of the af-FMC order in one of the four domains is reversed and there appear two corner spins directed to the diagonals of the square, which we call diagonally-directed (d-d) spins hereafter. 
In the $L_e=15$ (odd $L_e$) square, in which the af-FMC order having the 2$\times$2 periodicity mismatches with $L_e$, we have observed four states at lowest temperatures.
The ground state is characterized by two d-d spins on one of the diagonals as seen in Fig.~\ref{gr-st-0}(b-i), the lowest excited state by two d-d spins on one of the edges (b-ii), and the further excited state by no d-d spins but with a point defect (b-iii) or a dislocation (b-iv) of the af-FMC order inside the square.
In both even and odd squares, the difference between the total energies of the ground state and the low-lying excited state(s) is only of the order of $L^0$.

\begin{figure}[t]
\begin{center}
\includegraphics[width=0.6\linewidth]{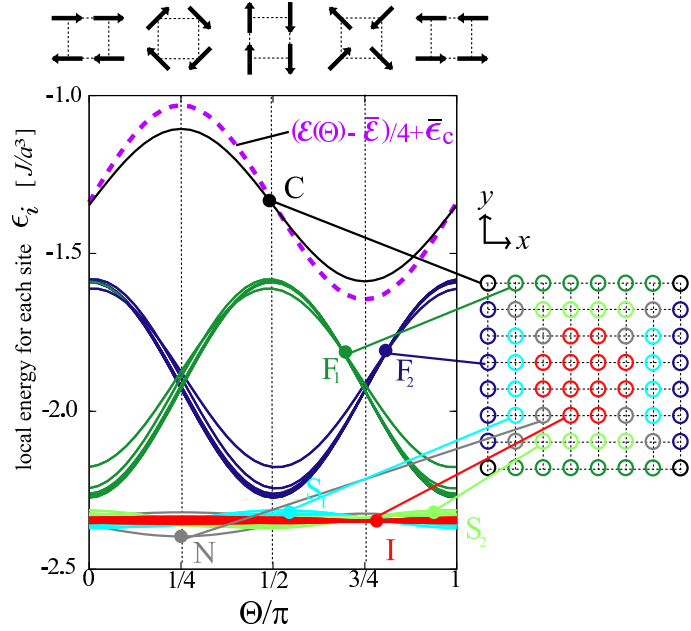}
\end{center}
\caption{The local energy at each site $i$, $\epsilon_i(\Theta)$, of the $\Theta$-LT state, which is a single domain state with the LT order of a fixed $\Theta$, of the $\Phi=0$ square with $L_e$=16. 
Curves F, S, C, N and I denote $\epsilon_i(\Theta)$ at sites on the first (green and  blue) and second (light green and light blue) lines from edges, at the  first (black) and the two next (gray) points along the diagonals, and at further interior sites (red). 
The $\Theta$-dependence of the total energy ${\cal E}(\Theta)$ divided by four is also shown by setting its average over $\Theta\ (=-143.42)$ at that of curve C, 
$\bar \epsilon_{\rm C}$ ($= -1.347$).}\label{Pot-0}
\end{figure}

In order to understand the energetics of the low-lying energy states simulated, it is easier for us to consider first a set of states, in which all spins on the square align as a single domain of the perfect LT order with a fixed value of $\Theta$. 
We call them the $\Theta$-LT states and denote their local energy at each site $i$ as $\epsilon_i(\Theta)$.
The sum of the latter over $i$ is twice of the total energy (Eq.(\ref{Ham})) of the $\Theta$-LT states and is denoted as ${\cal E}(\Theta)$.
All $\{ \epsilon_i(\Theta) \}$'s are shown in Fig.~\ref{Pot-0}. 
Reflecting the $O(2)$ symmetry of the LT order combined with the square geometry of the system, $\epsilon_i(\Theta)$ has the periodicity written as
\begin{equation}
\Delta\epsilon_i(\Theta) \equiv \epsilon_i(\Theta) - {\bar \epsilon}_i = \Delta\epsilon_i(\Theta+\pi),
\label{eq:d-epsilon}
\end{equation}
where ${\bar \epsilon}_i$ is the average of $\epsilon_i(\Theta)$ over $\Theta$.
An important observation in Fig.~\ref{Pot-0} is that $|\Delta\epsilon_i(\Theta)|$'s have significant magnitude only on edges (curves F$_1$ and F$_2$) and at corners (curve C).
For example, $|\Delta\epsilon_i(\Theta)|$'s at sites next to an edge (curves S$_1$ and S$_2$) are smaller than those of curves F$_1$ and F$_2$ by an order of magnitude.
Curves I at sites further inside become quickly independent of $\Theta$.
Moreover, since $\Delta\epsilon_i(\Theta)$'s on sites F$_1$ and F$_2$ nearly cancel with each other for a fixed $\Theta$, the $\Theta$-dependence of ${\cal E}(\Theta)$ is expected to come out dominantly from $\epsilon_i(\Theta)$'s at the four corners.
This is in fact the case as seen in Fig.~\ref{Pot-0} where we also draw the curve of $\Delta {\cal E}(\Theta)/4 = [{\cal E}(\Theta) - {\bar {\cal E}}]/4$.
Here ${\bar {\cal E}}$ is the average of ${\cal E}(\Theta)$ over $\Theta$.
We emphasize that ${\bar {\cal E}}/4 = -143.42$ whereas $|\Delta {\cal E}(\Theta)|/4 \simeq [|\Delta\epsilon_i(\Theta)|$ on the corner] $\simeq 0.24$ as shown in Fig.~\ref{Pot-0}.
The $\Theta$-LT state with minimum ${\cal E}(\Theta)$ is the micro-vortex state with $\Theta=3\pi/4$ whose four corner spins are in the vortex configuration, while that with maximum ${\cal E}(\Theta)$ is also the micro-vortex state but with $\Theta=\pi/4$ whose four corner spins are in pairs of the d-d spins.

\begin{figure}[t]
\begin{center}
\includegraphics[width=0.6\linewidth]{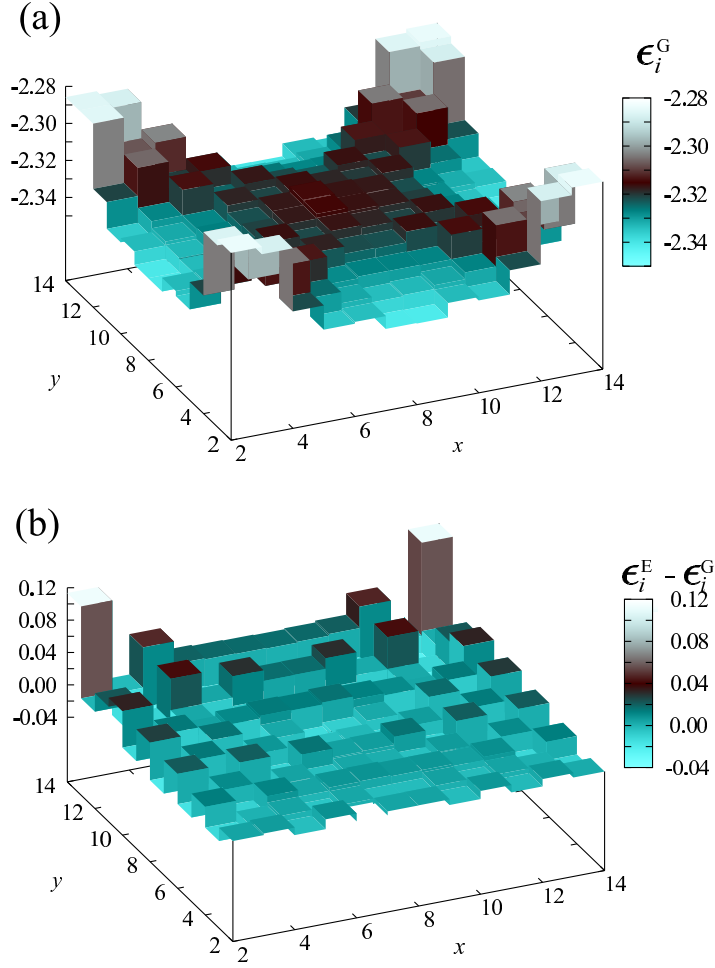}
\end{center}
\caption{The local energies of the simulated ground state $\{ { \epsilon^{\rm G}_i} \}$ of the $\Phi=0$ square with $L_e$=16 (a). 
Those of the boundary sites are omitted since they are out of the scale of the figure. 
The local energy differences of the excited state from the ground states, $\{ {\epsilon^{\rm E}_i - \epsilon^{\rm G}_i} \}$ (b). 
The difference at the two corners with the d-d spins ($\simeq 0.53$) is cut off.
}
\label{Pot-gr-st}
\end{figure}

As compared with the minimum ${\cal E}(\Theta)$ ($=-573.96$ with $\Theta=3\pi/4$), the ground state numerically simulated has the total energy, ${\cal E}^{\rm G} = -587.33$.
Its multi-domain structure is shown in Fig.~\ref{gr-st-0}(a-i). 
The latter is what we can simply think of from the above observation on the energetics of the $\Theta$-LT states.
All spins on the four edges align in parallel to each edge they lie on in the simulated ground state.
This yields energy gain relative to ${\cal E}(\pi/4)$ by the amount estimated as $\simeq 4 \times (L-2) \times [|\Delta\epsilon_i(\Theta)|$ of curve F$_1$ or F$_2$] $\simeq 18.51$.
This figure is 138$\%$ of the difference ${\cal E}(\pi/4) - {\cal E}^{\rm G}$.
Note that the ground-state energy gain from the four corner spins in the micro-vortex pattern is already taken into account by this energy difference.
We conclude therefore that the major gain of the ground-state energy relative to those of the $\Theta$-LT states is of the order of $L$, and can be attributed to the reduction of the symmetry of local energies in Eq.~(\ref{Ham}) from $O(2)$ in the bulk to Z$_2$ on the edges and corners.

To go one step further into the energetics of the ground state, let us examine fine details of its local energies, $\{ \epsilon_i^{\rm G} \}$.
For example, the value $\epsilon_i^{\rm G}$ of the corner spin is $-1.729$ as compared with $-1.588$ which is the minimum value of curve C in Fig.~\ref{Pot-0}, while the averaged value of $\epsilon_i^{\rm G}$'s of the edge spins is $-2.247$ as compared with $-2.252$ which is the minimum value of curve F$_1$ or F$_2$.
These fine details, however, give rise to only several percents modification on the above estimation of the ground-state energy gain.
The major part of the modification, or the excess energy of the ground state relatively to that of the $\Theta$-LT state comes out from $\epsilon_i^{\rm G}$'s of the interior spins which are shown in Fig.~\ref{Pot-gr-st}(a).
Note that, in the figure, the scale of its abscissa is enlarged and $\epsilon_i^{\rm G}$'s of the edge and corner spins, which are out of the scale of the abscissa, 
are omitted.
In each domains with either $x$- or $y$-directed af-FMC order, $\epsilon_i^{\rm G}$'s are nearly constant and their values are close to that of curve I in Fig.~\ref{Pot-0}.
On their boundaries, i.e., on the region of domain walls, $\epsilon_i^{\rm G}$'s are larger than those of spins within the domains. 
However, their differences are small and even their sum is at most about one fourth of the energy gains due to the boundary spins discussed above.
The circumstances are the same for spins near the center of the square where the four domain walls meet and the spin configuration is close to that of
the micro-vortex state (Fig.~\ref{gr-st-0}(a-i)).
Although we cannot define accurately the region of the domain walls, we can say that its width is significantly smaller than what we have in mind for domain walls in ordinary Heisenberg ferromagnets.

\begin{figure}[b]
\begin{center}
\includegraphics[width=0.6\linewidth]{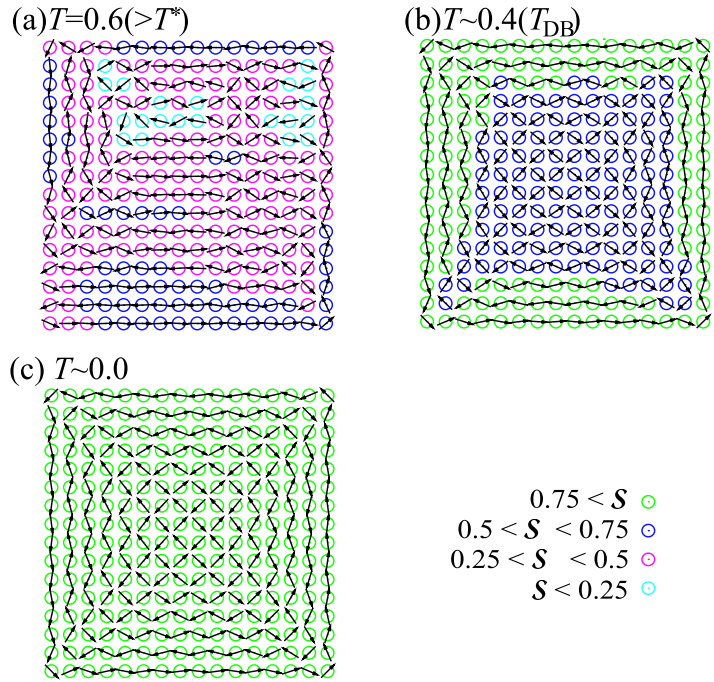}
\end{center}
\caption{Freezing spin configurations in one of the cooling runs of the 
 $\Phi=0$ $L_e=16$ dipolar square.}\label{freezP-0}
\end{figure}

We are faced to a similarly peculiar aspect of domain walls, when we examine the energetics of the excited state shown in Fig.~\ref{gr-st-0}(a-ii), where the af-FMC order of one of the domains in the ground state is reversed. 
In Fig.~\ref{Pot-gr-st}(b), we show differences $\{ \epsilon_i^{\rm E}-\epsilon_i^{\rm G} \}$, where $\epsilon_i^{\rm E}$ is the local energy at site $i$ in the excited state.
 The corresponding total excited energy is only 1.91, for which 1.06 arises from the two d-d spins at both ends of the reversed domain and 0.63 from the 18 sites neighboring to the corners.
The pattern of $\{ \epsilon_i^{\rm E}-\epsilon_i^{\rm G} \}$ at the remaining sites is rather flat.
This implies that the domain-walls both in the ground state and the excited states associate excess energies of an almost equal magnitude, and is naturally expected from the periodicity of $\epsilon_i(\Theta)$ represented by Eq.~(\ref{eq:d-epsilon}).
An important result here is that the excess energy of the excited state is almost confined at the two corners with d-d spins and at sites neighboring to them.

The spin configuration on the $L_e$=15 (odd $L_e$) dipolar square is intrinsically incompatible with the 2$\times$2 periodicity of the LT order.
Without going into details of the energetics of its low-lying states, we here only point out the following aspects.
If we remove spins, say, on the top and right edges from the ground-state configuration of the $L_e$=16 square (Fig.~\ref{gr-st-0}(a-i)), we obtain the ground state of the $L_e$=15 square shown in Fig.~\ref{gr-st-0}(b-i). 
The same procedure on the (a-ii) configuration yields the lowest excited state of (b-ii).
The boundary spins draw a closed loop in the next excited states but with an apparent point defect at the center as seen in (b-iii) and with an apparent line defect next to the right edge in (b-iv). 
Still, the total energies of the two states almost coincide with each other, and are higher than the ground-state energy of (b-i) by only 2.24, i.e., of the order of $L^0$. 
This is another observation on the peculiar nature of domain walls as well as defects and dislocations in a finite dipolar system.

Now let us discuss the freezing characteristics of the dipolar squares. 
We can see in Fig.~\ref{MLT-0} that the freezing branches are fixed, or thermally blocked, at around $T\simeq 0.4$ in the present simulation. 
We call this temperature as the domain-blocking temperature and denote it as $T_{\rm DB}$.  
In Fig.~\ref{freezP-0}, we show  the freezing patterns $\{{\cal S}_i\}$ observed in one of the cooling run of the $L_e=16$ square which reaches to the ground state. 
These figures lead us to the from-edge-to-interior freezing scenario~\cite{KSKTT} also for the present dipolar square.
Once the short-range LT order reaches the system size at around $T^*$, at which specific heat exhibits maximum (Fig.~\ref{SpeciC}), spins on the edges and their vicinity tend to align in an af-FMC order due to the symmetry reduction to Z$_2$ on the edges. 
As the temperature decreases further, each af-FMC domain grows, thereby it looks for a proper alignment with other domains  while it is overturning due to thermal fluctuation. 
There also appear and disappear domains with the micro-vortex LT order (Fig.~\ref{GS}(c)) inside the system.
The consequence is an almost common increase of the freezing parameter ${\cal S}$ and the order parameter $M_{\rm LT}$ as seen in Fig.~\ref{PV} and Fig.~\ref{MLT-0} (a), respectively. 
At around $T\simeq T_{\rm DB}$ domains (or spins on the edges) are thermally blocked.
At lower temperatures than $T_{\rm DB}$, ${\cal S}$ and $M_{\rm LT}$ further increase simply due to the reduction of thermal fluctuations on individual spins within the domains. 
But $M_{\rm LT}$ does not saturate to unity even at the lowest temperature simply because of the multi-domain structure of the ground state associated with the peculiar energetics discussed above.

\subsection{$\Phip4$ dipolar squares}

\begin{figure}[t]
\begin{center}
\includegraphics[width=0.6\linewidth]{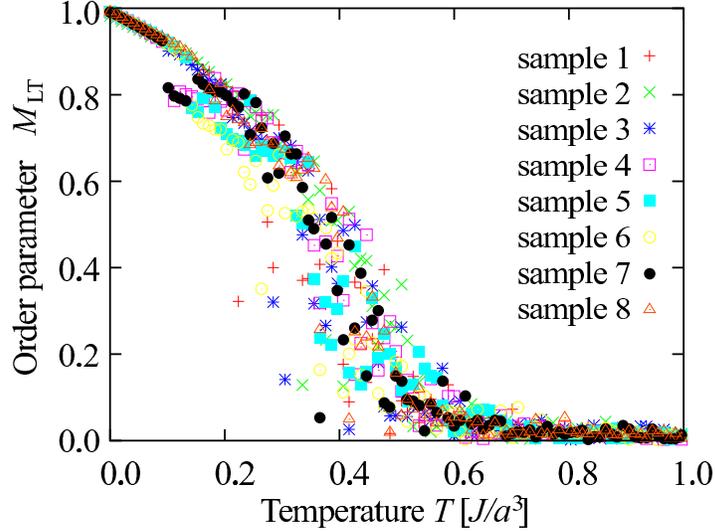}
\end{center}
\caption{The order parameter $M_{\rm LT}$ of $\Phip4$ squares with $L_d=22$ in each cooling run.}
\label{MLT-pi/4}
\end{figure}

\begin{figure}[t]
\begin{center}
\includegraphics[width=0.6\linewidth]{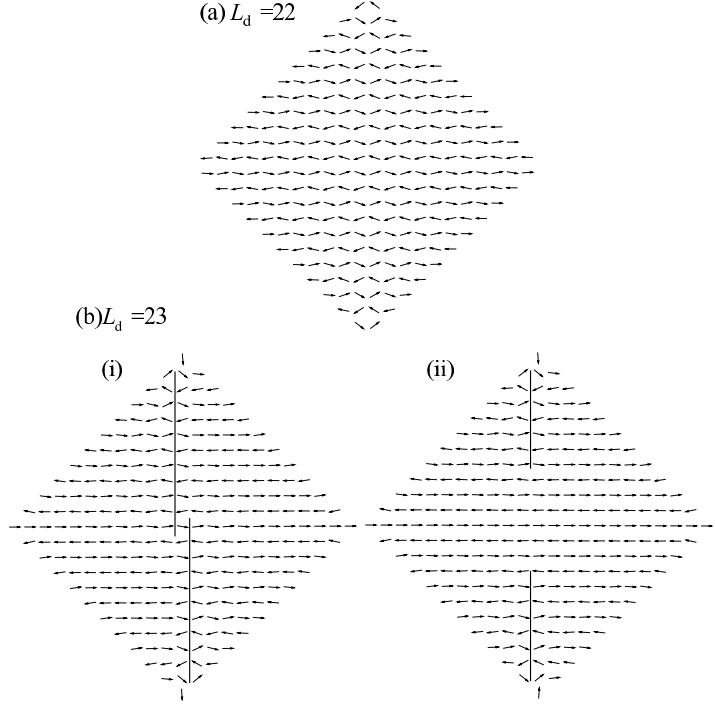}
\end{center}
\caption{Spin configurations at the lowest temperature obtained in 
the cooling processes shown in Fig.~\ref{MLT-pi/4}: (a) $L_d=22$ and (b) $L_d=23$. The solid line in (b) indicates a dislocation line.}\label{gr-st-pi/4}
\end{figure}

From the behavior of freezing parameter ${\cal S}$ shown in Fig.~\ref{PV}, one may expect that the freezing characteristic of the $\Phip4$ dipolar square with $L_d=22$ (even $L_d$) having a pair of spins at each corner and that of the $L_d=23$ (odd $L_d$) square having a single spin at each corner are not much different from each other. 
Actually this is also the case for $M_{\rm LT}$'s observed in eight independent cooling processes, the result of which is shown in Fig.~\ref{MLT-pi/4} only for the even-$L_d$ square.
In contrast to the $\Phi=0$ square, $M_{\rm LT}$'s of the $\Phip4$ square both with even- and odd-$L_d$ simulated end up with one branch at lowest temperatures and their values reach to unity within our numerical accuracy.
When, however, the spin configurations at the lowest temperature are looked at carefully, as shown in Fig.~\ref{gr-st-pi/4}, essentially only one pattern (a) is found in the even-$L_d$ square, while two are found for the odd-$L_d$ square (b-i,\ ii).
The three patterns consist of a single domain of the af-FMC order in parallel to one of the diagonals of the square (or one of the lattice axes) but with different modulations starting from both ends of the other diagonal. 

\begin{figure}[t]
\begin{center}
\includegraphics[width=0.5\linewidth]{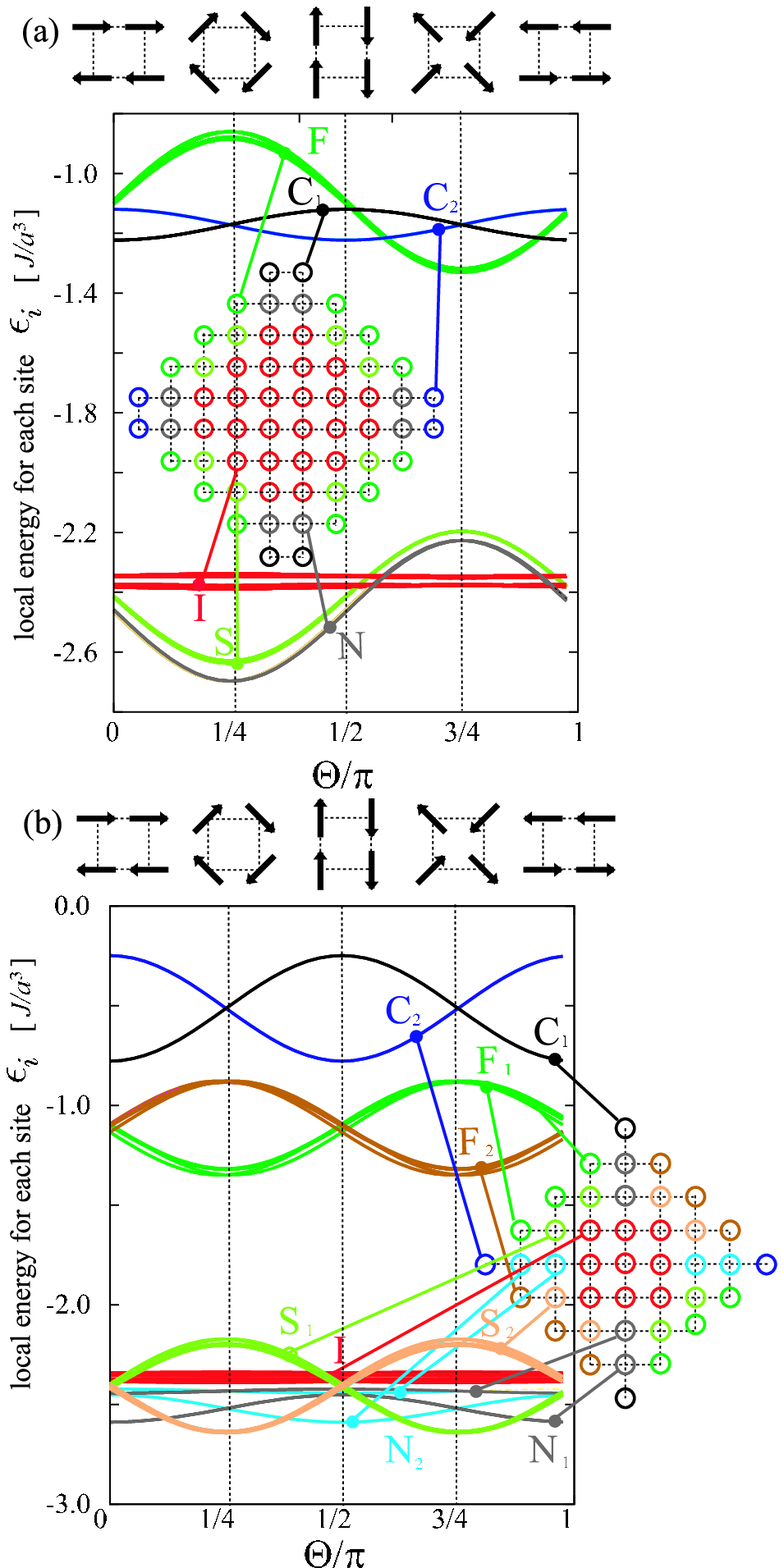}
\end{center}
\caption{The local energies $\epsilon_i(\Theta)$ of the $\Theta$-LT states of the $\Phi=\pi/4$, $L_d$=22 (a) and $L_d$=23 (b) squares. 
Curves F, S, C, N and I denote the energies at sites on the first (green and brown) and second lines (light green and fresh) from edges, at the corners (black and blue), at sites next to the corners (gray and light blue), and at the interior sites (red). 
Note that in the odd-$L_d$ square, curves N$_1$ and N$_2$ are in phase with curves C$_1$ and C$_2$, respectively.
}\label{Pot}
\end{figure}

Let us first introduce the $\Theta$-LT states and examine their energetics as in the previous subsection.
Figure~\ref{Pot}(a) shows $\epsilon_i(\Theta)$'s of the even-$L_d$ square.
One can see clearly the short-ranged characteristics of $\Delta\epsilon_i(\Theta)$'s also in the present square, namely, $\Delta\epsilon_i(\Theta)$'s at interior sites only a few lattice distances from the boundary become almost independent of $\Theta$.
One of the characteristic aspects of the figure is that $\Delta\epsilon_i(\Theta)$ at edge sites (curves F) and at sites next to the edges (curves S) have almost equal amplitudes and are almost perfectly out-of phase to each others. 
This implies that their sum (adding $\Delta\epsilon_i(\Theta)$ at one of the two N sites to the sum), which we may regard as the sum of $\Delta\epsilon_i(\Theta)$'s at sites on a zigzag edge, becomes independent of $\Theta$.
The consequence is the absence of energy gain of the order of $L$ from each zigzag edge, in a sharp contrast to the geometry effect in the $\Phi=0$ square. 
Another aspect to be noted is that there appear two types of $\epsilon_i(\Theta)$ at the corner sites, curves C$_1$ and C$_2$, whose $\Delta\epsilon_i(\Theta)$'s have also equal amplitudes and are perfectly out-of phase to each others.
Reflecting these aspects due to the geometry of the present dipolar square, the total energy of the $\Theta$-LT state turns out to be almost independent of $\Theta$, i.e., ${\bar {\cal E}}= -569.54$ with $|\Delta {\cal E}(\Theta)| < 0.86$.
The above arguments can be straightforwardly extended to the energetics of the odd-$L_d$ square, by taking into account the 
further fine detail
that both curves F and S are split into two types as shown in Fig.~\ref{Pot}(b).
We obtain ${\bar {\cal E}}=-569.33$ with $|\Delta {\cal E}(\Theta)| \sim 0.00$ for this dipolar square.

\begin{figure}[t]
\begin{center}
\includegraphics[width=0.6\linewidth]{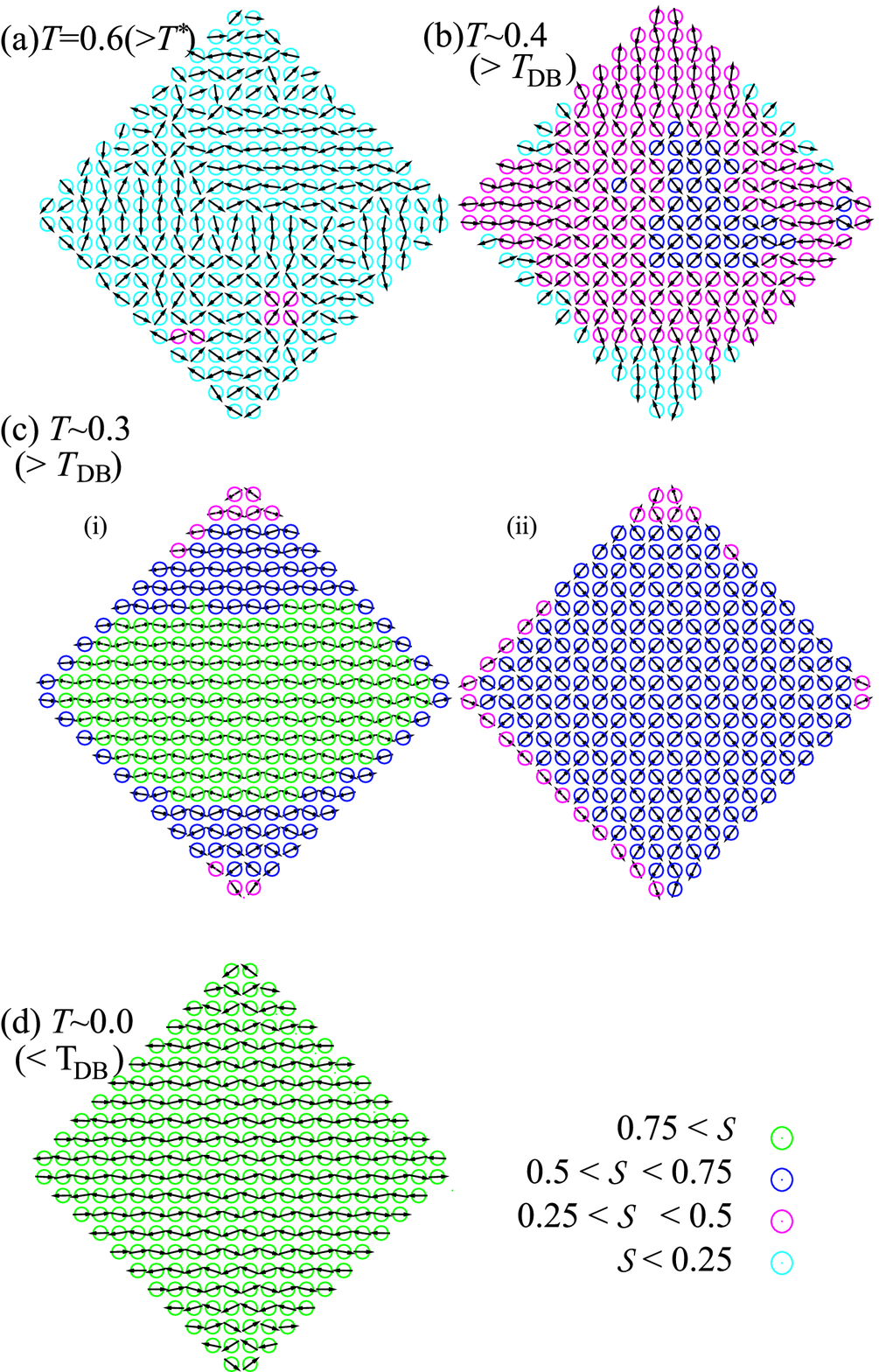}
\end{center}
\caption{The directions of the spins on the $x$-$y$ plane and the local freezing parameter ${\cal S}_i$ (a) at $T$=0.6, (b) 0.4, (c) 0.3, (d) 0.0 for $\Phi=\pi/4$ ($L_d$=22).}\label{le-pi/4}
\end{figure}

The total energies of the simulated ground state are lower than ${\bar {\cal E}}$ above described by 2.07 and 2.17 for the even- and odd-$L_d$ squares, respectively.
As seen in Fig.~\ref{gr-st-pi/4}(a), the ground state of the even-$L_d$ square consists of a single domain with the af-FMC order directed in parallel to one of the diagonals of the square with dislocations of the $\Theta=\pi/4$ type which start from both ends of the other diagonal and go deeply into the interior almost up to the center of the square where the two dislocations meet together smoothly.
At and near these two ends where the micro-vortex units appear, the ground state gains energy, while it looses energy at and near the dislocations inside the square relatively to $\epsilon_i(\Theta=0)$.
The situation is similar but a little complicated for the odd-$L_d$ square. 
In its ground state (Fig.~\ref{gr-st-pi/4}(b-i)), there appear two d-d spins at two corners of the other diagonal.
The dislocations of the $\Theta=\pi/4$ type start from sites next to the corners and go deeply into the interior almost up to the center of the square where the two dislocations meet together but are separated by one lattice distance in this case.
Our cooling simulations on this square end up also with the excited state shown in Fig.~\ref{gr-st-pi/4}(b-ii).
In the excited state, the two d-d spins are antiparallel to each other.
The centers of the two dislocations lie on an identical lattice axis, and the two dislocations are separated significantly with each others by the region with the complete af-FMC order.
We note that the excess energy of the excited state from the ground-state energy is only 0.30. 
Although we can hardly explain these energy gains and losses with enough quantitative accuracy, we consider that they represent the nature of the stiffness energy in the spin order due to the dipole-dipole interaction,~\cite{DBell} and will be discussed elsewhere.

We next argue the freezing property of the even-$L_d$ square demonstrated in Fig.~\ref{le-pi/4}.
At $T=0.6$, which is a little higher than $T^*$, we can see many short-ranged domains of various LT orders, sizes of which are a few tenths of $L$ (a). 
As the temperature decreases, the sizes of the domains grow (b), and at such low temperatures as $T \sim 0.3$, almost a single domain state of the af-FMC order (c-i) as well as that of the micro-vortex state (c-ii) are observed.
But the pattern is still fluctuating among such nearly single-domain configurations (mostly triggered by the discrete, though small, temperature decreases). 
This corresponds to the two branches of $M_{\rm LT}$ at temperatures from 0.1 to 0.3 in Fig.~\ref{MLT-pi/4}.
The upper and lower branches correspond to the states shown in Figs.~\ref{le-pi/4}(c-i) and (c-ii), respectively.
We specify the domain blocking temperature of this dipolar square as $T_{\rm DB} \simeq 0.1$ where the lower branch disappears.
Below $T_{\rm DB}$, the present MD simulation ends up always with the ground state pattern of (d), or Fig.~\ref{gr-st-pi/4}(a).
We interpret this result as follows.
Between the nearly $\Theta=0$ state of Fig.~\ref{le-pi/4}(c-i) and $\Theta=3\pi/4$ state of (c-ii), there exists a free energy barrier whose height is very limited, or of a comparable magnitude with the energy difference of the two states itself. 
The latter is expected to be also small ($\nle 1.0$) by judging from the energy difference between the corresponding $\Theta$-LT states.
The result also implies the absence of other low-lying excited states involving such extended defects as dislocations in this system.
The freezing process of the odd-$L_d$ square (not shown) is almost the same with that of the even-$L_d$ square at higher temperatures than $T_{\rm DB} \ (\nle 0.1)$.
In contrast to the even-$L_d$ square, however, our cooling simulations on the odd-$L_d$ square end up also with the excited state of Fig.~\ref{gr-st-pi/4}(b-ii) at the lowest temperature, implying that the excited state is separated from the ground state of (b-i) by a free energy barrier of a height significantly higher than their energy difference, ${\bar {\cal E}} - {\cal E}^{\rm G} \simeq 2.1$.
These minute but very peculiar aspects of the free energy structure are, we consider, the consequence of the intrinsic frustration effect of the dipole-dipole interaction combined with the geometry effect of these dipolar squares.
The proper understanding of its details is remained as a future problem.

\section{Summary and Discussion}

By the MD simulation based on the LLG equation at finite temperatures, we have investigated the magnetic ordering and freezing properties of the simplest finite dipolar systems, i.e., the dipolar squares cut out from a square lattice. 
Distinctly different phenomena are found depending on how a square is cut out (geometry effect).
The ground state of the $\Phi=0$ square, whose edges are parallel to the lattice axes, is found to have a multi-domain structure which consists of domains having the two types of the bulk LT (af-FMC) order. 
The peculiar ``from-edge-to-interior freezing'' to the ground state is observed as in the dipolar cube.~\cite{KSKTT}
These results are attributed to the reduction of the symmetry of the dipole-dipole interaction energy from $O(2)$ in the bulk to Z$_2$ on the edges. 
The symmetry reduction is, in turn, attributed to the peculiar anisotropic nature of the dipole-dipole interaction that it depends on the relative direction of sites in a real space of the interacting spins.
In the $\Phi=\pi/4$ square, whose edges are rotated by $\pi/4$ from those of the $\Phi=0$ square, the effect of the same symmetry reduction as above is canceled in the local energies of each of the four zigzag edges of the square. 
Its ground state consists of a single domain with the af-FMC order in parallel to the one of the diagonals of the square with modulations starting from both ends of the other diagonal.
The freezing starts from the interior, and the final thermal blocking to a state at lowest temperatures is governed by spin configurations at and near both ends of the other diagonal.

The excess energy associated with a domain wall in the multi-domain ground state of the $\Phi=0$ square is evaluated to be of the order of $L^0$ with $L$ being a linear dimension of the square. 
This is similar to that of a domain wall in an ordinal Heisenberg ferromagnetic square under the antiperiodic boundary condition.
In contrast to the latter domain wall whose width is of the order $L$, however, 
that of the present dipolar square seems rather small though we cannot specify it accurately. 
In the ground state of the $\Phi=\pi/4$ square, we observe dislocations in the af-FMC order directed in parallel to one of the diagonals of the square.
They start from both ends of the other diagonal and go deeply into the interior. 
Still, their associated excess energy is of the order of $L^0$ and their width looks rather small, as is the case for the domain wall in the $\Phi=0$ square.
We consider that these characteristics of apparently extended objects, domain walls and dislocations, reflect the intrinsic frustration effect of the dipole-dipole interaction.
The frustration effect, which yields the peculiar ground state with the continuous $O(2)$ symmetry already in the bulk square lattice, makes it possible for spins to adjust to further (geometrical) restriction imposed in a finite system only with a quite small excess energy.

In the pattern of the local magnetic energies of the ground states, $\{ \epsilon_i^{\rm G} \}$, its major changes induced by the presence of the boundary are seen to propagate to the interior by only one or two lattice distances from the boundary.
Their spatial dependence associated with the domain walls and dislocations mentioned above is quite small in the sense that the total excess energy of an excited state associated with such a defect is only of the order of $L^{0}$. 
It is rather surprising that the long-ranged nature of the dipole-dipole interaction seems not to play an essential role on determining the ground-state and excited-state configurations. 
To get a deeper insight on roles of the long-ranged nature of the interaction, we have preliminarily carried out a similar analysis to that we adopted in our previous study on the dipolar cube,~\cite{KSKTT} namely, repeating the identical simulations but with the dipole-dipole interaction whose range is artificially cut off, and comparing the results of the two simulations. 
The consequence of the $\Phi=0$ square is that the long-ranged nature makes easier for the system to reach the ground-state multi-domain structure as observed in our previous work.
For the $\Phi=\pi/4$ square, on the other hand, the clear consequence has not been obtained yet because of the small energy differences between the states with and without dislocations. 
The problem remains for a future study.

Lastly we append an explanatory note on our MD simulation.
Most of the freezing characteristics reported in the present paper is rather robust to the cooling rate of the simulation under a fixed value of the temperature decrease $\Delta T$.
For example, a rapid cooling with $\tau$ one order smaller than the present one does not yield a significant change either in the peak temperature of the specific heat $T^*$ or the thermal blocking temperature $T_{\rm DB}$.  
In this sense we consider that our choice of the parameters $\tau$ and $\Delta T$ is appropriate for extracting the characteristic properties of the ground state and the low-lying excited states, as well as the freezing characteristics of finite dipolar squares of our interest. 
From a technological point of view on nanomagnetism, the value of $T_{\rm DB}$ of a given dipolar square (whether it is higher than the room temperature or not) is certainly of importance. 
To answer this question, we need to know, not only the systematic
 $L$-dependence of the energy difference between the ground and excited states that we have investigated in the present work, but also that of the free energy barrier height between the states.
But such an analysis is beyond the scope of the present work.

\section*{Acknowledgment}

The present work is supported by the Next Generation Supercomputing Project, Nanoscience Program, MEXT, Japan.
The numerical simulations have been partially performed also by using the facilities at the Supercomputer Center, Institute for Solid State Physics, the University of Tokyo.

\end{document}